\documentclass[aps,prl,twocolumn,superscriptaddress]{revtex4}
\usepackage{graphicx}
\usepackage{dcolumn}
\usepackage{bm}
\usepackage{wrapfig}
\usepackage{amssymb}
\usepackage{amsmath, amsthm}
\usepackage{color}
\usepackage{multirow}
\usepackage{pdfsync}
\usepackage{cancel}
\usepackage[normalem]{ulem}

\def\be{\begin{equation}}
\def\ee{\end{equation}}
\def\bea{\begin{eqnarray}}
\def\eea{\end{eqnarray}}
\def\la{\langle}
\def\ra{\rangle}
\definecolor{darkgreen}{rgb}{0, 0.4, 0}

\newcommand{\ie}{{\it{i.e.~}}}

\newtheorem{theorem}{Theorem}
\newtheorem{lemma}{Lemma}
\newtheorem*{definition*}{Definition}

\def\x{\mathbf{x}}
\def\a{\mathbf{a}}

\begin{document}

\title{Can observed randomness be certified to be fully intrinsic?}

%
%
%

\author{Chirag Dhara$^{1,*}$, Gonzalo de la Torre$^{1,*}$, Antonio Ac\'\i n$^{1,2}$\\[0.5em]
{\it\small $^1$ICFO--Institut de Ciencies Fotoniques, E--08860 Castelldefels, Barcelona, Spain}\\
{\it\small $^2$ICREA--Institucio Catalana de Recerca i Estudis
Avan\c{c}ats, E--08010 Barcelona, Spain}\\
{\small $^*$(These authors contributed equally to this work.)}}

\date{\today}

\begin{abstract}
Randomness comes in two qualitatively different forms.  Apparent
randomness can result both from ignorance or lack of
control of degrees of freedom in the system. In contrast,
intrinsic randomness should not be ascribable to any such cause.
While classical systems only possess the first kind of randomness,
quantum systems are believed to exhibit some intrinsic randomness.
In general, any observed random process includes both forms of
randomness. In this work, we provide quantum processes in
which all the observed randomness is fully intrinsic. These
results are derived under minimal assumptions: the validity of the
no-signalling principle and an arbitrary (but not absolute) lack
of freedom of choice. The observed randomness tends to a perfect
random bit when increasing the number of parties, thus defining an
explicit process attaining full randomness amplification.
\end{abstract}

\maketitle

Physical theories aim at providing the best possible predictions
for the phenomena occurring in nature. Consequently, on observing
a probabilistic process, a natural question arises: how much -if
any- of that is intrinsically unpredictable?

Consider an experimental setup in which a variable takes different
values with different probabilities. This variable has
\emph{observed randomness} that can be easily estimated from the
measured statistics. In general, we can distinguish two
qualitatively different forms of randomness contributing to the
observed randomness of a process. The first is the \emph{apparent
randomness}, which appears as a consequence of
imperfections of the system, such as lack of knowledge and control
of all the relevant degrees of freedom. Clearly, an improvement
on our control of the setup reduces this form of randomness. The
second form of randomness is termed \emph{intrinsic randomness}
and refers to the component of observed randomness that cannot be
ascribed to imperfections. It is this second form of randomness
that should be considered truly random, as any improvement on our
control of the setup leaves it unchanged.

The quantitative contribution of each form of randomness to the
observed randomness depends on the physical theory used to
describe the process. In classical theories, for instance, all
observed randomness is apparent, as it is always possible to
explain any random classical process as the probabilistic mixture
of deterministic classical processes~\cite{Bell1964,Fine1982a}.
Moving to the quantum domain, the axioms of quantum theory state
that measurements on quantum particles yield intrinsically random
outcomes. Yet, the fact that a theory makes predictions only in
terms of probabilities does not necessarily imply the existence of
intrinsic randomness. It may simply reflect some limitations of
the formalism, in the sense that a better, more complete theory
could restore determinism~\cite{Einstein1935, Bohr1935}.
However, the non-local correlations observed when measuring
entangled particles allow one to assess the randomness of a
process independent of the full quantum formalism. Under only two
assumptions, (i) the impossibility of instantaneous communication,
- known as the no-signaling principle 
- and (ii) that the measurement settings in a Bell test can be
chosen at random - known as freedom of choice - non-local quantum
correlations necessarily imply intrinsic randomness \cite{Cavalcanti2012}. This is
because such correlations cannot be described as the probabilistic
mixture of \textit{deterministic} processes.

Up to now, in all Bell tests, the intrinsic randomness revealed by
quantum non-locality (under said assumptions) is also
mixed with apparent randomness, resulting from the
non-completeness of quantum theory. In this work, we ask the
following fundamental question: is there any quantum process that
is as intrinsically random as it is observed to be? We answer this
question in the affirmative by providing a family of
quantum processes whose intrinsic randomness can be computed analytically for arbitrary system sizes and also demonstrating that this is strictly equal to the observed randomness.

Our results are related to recent attempts to prove the
completeness of quantum physics. In~\cite{Colbeck2011a}, Colbeck
and Renner claimed that no no-signalling theory can have a
better predictive power than quantum theory. However, the proof,
which is based on the quantum violation of the chained Bell
inequality, assumes that the settings in the inequality can be
chosen freely. This introduces a circularity in the argument, as
the free process needed in the proof is already assumed to be
complete. A possible way out to break this circularity is to
consider protocols for randomness
amplification~\cite{Colbeck2012}. There, the intrinsic randomness
of a quantum process could be proven using a source of imperfect
randomness. In fact, the protocol for full
randomness amplification given in~\cite{Gallego2012a} provides a
Bell test in which a measured variable has an intrinsic randomness
that tends to be equal to the observed randomness in the limit of
an infinite number of parties. Our results provide finite-size
Bell setups in which observed and intrinsic randomness are
\emph{strictly} equal using \emph{arbitrarily small} randomness
for the choice of measurements. In this sense, they represent the
strongest proof of completeness of a quantum process.

{\it Preliminaries.---} Suppose that a Bell test is performed
repeatedly among $N$ parties and the resulting statistics is given
by $P_{\rm obs}(\a|\x)$, where $\a=(a_1,\ldots,a_N)$ and
$\x=(x_1,\ldots,x_N)$ are the string of outcomes and measurement
inputs of the parties involved. Let $g$ be a function acting on
the measurement results $\a$. As previously explained, there are
different physically relevant notions of randomness.

First, the \textit{observed randomness} of $g$ for measurements
$\x$ is the randomness computed directly from the statistics.
Operationally, this may be defined as the optimal probability of
guessing the outcome of $g$ for input $\x$,
\begin{equation}
\label{observedrandomness}
G_{\rm obs}(g,\x,P_{\rm obs})=\operatorname*{max}_{k\in {\rm Im}(g)}~P_{\rm obs}(g(\a)=k|\x).
\end{equation}
where ${\rm Im}(g)$ is the image of function $g$.

Moving to the definition of the \emph{intrinsic randomness}, one
should consider all possible preparations of the observed
statistics in terms of no-signalling probability distributions. In
our context, a particular preparation reads
\begin{equation}
\label{preparation} P_{\rm obs}(\a|\x)=\sum_e p(e|\x) P_e^{\rm
ex}(\a|\x)
\end{equation}
where the $P_e^{\rm ex}$ are extremal points of the no-signaling
set~\cite{Barrett2005b}. The terms $p(e|\x)$ may depend on $\x$,
which accounts for possible correlations between the preparation
$e$ and the measurement settings $\x$, given that the choice of
measurements are not assumed to be free.
Hence, we define the intrinsic randomness of a function $g$ by
optimizing over all possible non-signalling preparations of
$P_{\rm obs}$ so as to minimize the randomness of $g$.
In other words,
\begin{align}
& G_{\rm int}(g,\x,P_{\rm obs})= \max_{\{p(e|\x),P_e^{\rm ex}\} } \sum_{e} p(e|\x) G_{\rm obs}(g,\x,P_e^{\rm ex})\nonumber \\
 &\text{subject to:} \nonumber\\
&\sum_e p(e|\x) P_e^{\rm ex}(\a|\x)=P_{\rm obs}(\a|\x)  \label{sumupto} \\
&p(\x|e)\geq \delta ~~\text{with}~~ \delta>0; ~\forall ~\x, e
\label{relaxfchoice}
\end{align}
where $G_{\rm obs}(g,\x,P_e^{\rm ex})=\max_{k}~P_e^{\rm
ex}(g(\a)=k|\x)$ is  also the intrinsic randomness of $P_e^{\rm
ex}$, since intrinsic and observed randomness must coincide for
extremal points of the non-signalling set. Note that
condition $p(\x|e)\geq \delta>0$ allows for an arbitrary (but not
absolute) relaxation of the freedom of choice assumption by
allowing for arbitrary (yet not complete) correlations between the
preparation and the measurement settings. Physically, this
condition ensures that all measurement combinations appear for all
possible preparations $e$. An example of a source of randomness
fulfilling this condition is a Santha-Vazirani source
\cite{Santha1986}. Note however that our definition allows sources
more general than the Santha-Vazirani sources.


From a cryptographic point of view, the observed randomness is the
one perceived by the parties performing the Bell test, whereas the
intrinsic randomness is that perceived by a non-signalling
eavesdropper possessing knowledge of the preparation of the
observed correlations and with the ability to arbitrarily (yet not
fully) bias the choice of the measurement settings.


In general, $G_{\rm obs}$ is strictly larger than $G_{\rm intr}$,
as the set of non-signalling correlations is larger than the
quantum. The results in~\cite{Barrett2006,Colbeck2012a} provide a
Bell test in which $G_{\rm intr}$ approaches $G_{\rm obs}$ (and to
$1/2$) in the limit of an infinite number of measurements and
assuming free choices, that is, $p(\x|e)$ in \eqref{preparation}
is independent of $e$. The results in~\cite{Colbeck2012} allow
some relaxation of this last condition. The results
in~\cite{Gallego2012a} arbitrarily relaxed the free-choice
condition and give a Bell test in which $G_{\rm intr}$ tends to
$G_{\rm obs}$ (and both tend to $1/2$) in the limit of an infinite
number of parties. Here, we provide a significantly stronger
proof, as we allow the same level of relaxation on free choices
and provide Bell tests in which $G_{\rm intr}=G_{\rm obs}$ for any
number of parties. Moreover, a perfect random bit is obtained in
the limit of an infinite number of parties.

{\it Scenario.---}
Our scenario consists of $N$ parties where each performs two
measurements of two outcomes. In what follows, we adopt a
spin-like notation and label the outputs by $\pm 1$. Then, any
non-signalling probability distribution can be written as (for
simplicity we give the expression for three parties, but it easily
generalizes to an arbitrary number)
\begin{equation}
\label{pns}
\begin{aligned}
P(&a_1,a_2,a_3|x_1,x_2,x_3) =\\
&\frac{1}{8}\left(1+a_1\la A^{(x_1)}_1\ra +a_2 \la A^{(x_2)}_2\ra
+a_3\la
A^{(x_3)}_3\ra +\right.\\
&\left.a_1a_2 \la A^{(x_1)}_1A^{(x_2)}_2\ra+a_1a_3\la
A^{(x_1)}_1A^{(x_3)}_3\ra \right.+\\
&\left.a_2a_3 \la A^{(x_2)}_2A^{(x_3)}_3\ra+a_1a_2a_3\la
A^{(x_1)}_1A^{(x_2)}_2A^{(x_3)}_3\ra\right) ,
\end{aligned}
\end{equation}
where $A^{(x_i)}_i$ denotes the outputs of measurement $x_i$ by each
party $i$. In this scenario, we consider Mermin Bell inequalities,
whose Bell operator reads \be \label{mermindefinition}
M_N=\frac{1}{2}
M_{N-1}(A^{(0)}_N+A^{(1)}_N)+\frac{1}{2}M_{N-1}'(A^{(0)}_N-A^{(1)}_N),
\ee where $M_2$ is the Clauser-Horne-Shimony-Holt operator and
$M_{N-1}'$ is obtained from $M_{N-1}$ after swapping
$A^{(0)}_i\leftrightarrow A^{(1)}_i$. We study probability
distributions that give the maximal non-signalling violation of
the Mermin inequalities and focus our analysis on a function $f$
that maps the $N$ measurement results into one bit as follows:
\begin{equation}
\label{fdefined}
\begin{aligned}
f(\a)= \left\{
\begin{array}{c l}
+1~~ & n_-(\a)=(4j+2) ;~~{\rm with~} j\in \{0,1,2,\ldots\} \\
\\
-1~ & \text{otherwise}
\end{array}
\right .
\end{aligned}
\end{equation}
where $n_-(\a)$ denotes the number of results in $\a$ that are
equal to $-1$.

{\it Results.---} Our goal in what follows is to quantify the
intrinsic randomness of the bit defined by $f(\a)$ for those
distributions maximally violating the Mermin inequality for odd
$N$. We first prove the following
\begin{lemma}
\label{Theorem2principal} Let $P_{\rm M}(\a|\x)$ be an $N$-partite
(odd $N$) non-signalling probability distribution maximally
violating the corresponding Mermin inequality. Then, for any input
$\x$ appearing in the inequality \be \label{property}
P_{\rm M}(f(\a)=h_N|\x)\geq 1/2, ~{\rm with~} h_N=\sqrt{2}\cos\left (\frac{\pi (N+4)}{4} \right ).\ee
\end{lemma}


Note that, as $N$ is odd, $h_N=\pm 1$. Operationally, the Lemma
implies that, for all points maximally violating the Mermin
inequality, the bit defined by $f$ is biased towards the same
value $h_N$. Since the proof of the Lemma for arbitrary odd $N$ is
convoluted, we give the explicit proof for $N=3$ here, which
already conveys the main ingredients of the general proof, and
relegate the generalization to the Appendix.

{\it Proof for three parties.---} With some abuse of notation, the
tripartite Mermin inequality may be expressed as, \be
\label{mermin3} M_3=\la 001\ra  + \la 010\ra  + \la 100\ra  - \la
111\ra  \leq 2, \ee where $\la x_1x_2x_3\ra = \la
A^{(x1)}_1A^{(x2)}_2A^{(x3)}_3\ra$  and similar for the other terms.
The maximal non-signalling violation assigns $M_3=4$ which can
only occur when the first three correlators in \eqref{mermin3}
take their maximum value of $+1$ and the last takes its minimum of
$-1$.

Take any input combination appearing in the
inequality~\eqref{mermin3}, say, $\mathbf{x_m}=(0,0,1)$. Maximal
violation of $M_3$ imposes the following conditions:
\begin{itemize}

\item[1.] $\la 001\ra  =1$. This further implies $\la 0\ra_1=\la 01\ra_{23} $, $\la 0\ra_2=\la 01\ra_{13}$ and $\la 1\ra_3=\la 00\ra_{12}$.
\item[2.] $\la 010\ra  =1$ implying $\la 0\ra_1=\la 10\ra_{23} $, $\la 1\ra_2=\la 00\ra_{13}$ and $\la 0\ra_3=\la 01\ra_{12}$.
\item[3.] $\la 100\ra  =1$ implying $\la 1\ra_1=\la 00\ra_{23} $, $\la 0\ra_2=\la 10\ra_{13}$ and $\la 0\ra_3=\la 10\ra_{12}$.
\item[4.] $\la 111\ra  =-1$ implying $\la 1\ra_1=-\la 11\ra_{23} $, $\la 1\ra_2=-\la 11\ra_{13}$ and $\la 1\ra_3=-\la 11\ra_{12}$

\end{itemize}
Imposing these relations on \eqref{pns} for input
$\mathbf{x_m}=(0,0,1)$ one gets
\begin{equation}
\begin{aligned}
\label{eqn10} P_{\rm
M}(a_1,a_2,a_3&|0,0,1)=\frac{1}{8}\left(1+a_1a_2a_3
+(a_1+a_2a_3)\la 0\ra_1+\right.\\
&\left.(a_2+a_1a_3)\la 0\ra_2+(a_3+a_1a_2)\la 1\ra_3\right)
\end{aligned}
\end{equation}

Using all these constraints and the definition of the function
\eqref{fdefined}, Eq.~\eqref{property} can be expressed as
\begin{eqnarray}
\begin{aligned}
&P_M(f(\a)=+1|\mathbf{x_m}) \\
& = P_{\rm M}(1,-1,-1|\mathbf{x_m})+P_{\rm
M}(-1,1,-1|\mathbf{x_m})\\
&+P_{\rm
M}(-1,-1,1|\mathbf{x_m}) \\
&= \frac{1}{4}(3-\la 0\ra_1 - \la 0\ra_2 -\la 1\ra_3)
\end{aligned}
\end{eqnarray}
Proving that $P(f(\a)=+1|\mathbf{x_m})\geq 1/2$ then amounts to
showing that $\la 0\ra_1 + \la 0\ra_2 +\la 1\ra_3\leq 1$. This
form is very convenient since it reminds one of a positivity
condition of probabilities.

We then consider the input combination $\mathbf{\bar{x}_m}$ such
that all the bits in $\mathbf{\bar{x}_m}$ are different from those
in $\x_m$. We call this the swapped input, which in the previous
case is $\mathbf{\bar{x}_m}=(1,1,0)$. Note that this is
\textit{not} an input appearing in the Mermin inequality. However,
using the previous constraints derived for distributions $P_{\rm
M}$ maximally violating the inequality, one has
\begin{equation}
\begin{aligned}
&P_{\rm
M}(a_1,a_2,a_3|1,1,0)\\
=&\frac{1}{8}(1+a_1\la 1\ra_1 +a_2 \la 1\ra_2 +a_3\la 0\ra_3 +a_1a_2 \la 11\ra_{12}\\ &+a_1a_3\la 10\ra_{13} +a_2a_3 \la 10\ra_{23}+a_1a_2a_3\la 110\ra_{123})\\
=&\frac{1}{8}(1+a_1\la 1\ra_1 +a_2 \la 1\ra_2 +a_3\la 0\ra_3 -a_1a_2 \la
1\ra_{3}\\ &+a_1a_3\la 0\ra_{2} +a_2a_3 \la 0\ra_{1}+a_1a_2a_3\la 110\ra_{123}),
\end{aligned}
\end{equation}
where the second equality results from the relations $\la
11\ra_{12}=-\la 1\ra_3$, $\la 10\ra_{13}=\la 0\ra_2$ and $\la
10\ra_{23}=\la 0\ra_1$.

It can be easily verified that summing the two positivity
conditions $P_{\rm M}(1,1,-1|\mathbf{\bar{x}_m})\geq 0$ and
$P_{\rm M}(-1,-1,1|\mathbf{\bar{x}_m})\geq 0$ gives the result we
seek, namely $1-\la 0\ra_1 - \la 0\ra_2 -\la 1\ra_3\geq 0$, which
completes the proof. $\Box$

Using the previous Lemma, it is rather easy to prove the following

\begin{theorem}
\label{Theorem1principal} Let $P_{\rm obs}(\a|\x)$ be an
$N$-partite (odd $N$) non-signalling probability distribution
maximally violating the corresponding Mermin inequality. Then the
intrinsic and the observed randomness of the function $f$ are
equal for any input $\x$ appearing in the Mermin inequality:
\[G_{\rm int}(f,\x,P_{\rm obs})=G_{\rm obs}(f,\x,P_{\rm obs})\]

where
\begin{eqnarray*}
G_{\rm obs}(f,\x,P_{\rm obs})&=&\max_{k\in \{+1,-1\}}P_{\rm obs}(f(\a)=k|\x)  \\
\end{eqnarray*}

\end{theorem}

{\it Proof of Theorem 1.---} Since $P_{\rm obs}$ maximally and
algebraically violates the Mermin inequality, all the extremal
distributions $P_e^{\rm ex}$ appearing in its decomposition must
also necessarily lead to the maximal violation of the Mermin
inequality (see Appendix for details). Hence, the
randomness of $f$ in these distributions as well satisfies Eqn.
\eqref{property} of Lemma \ref{Theorem2principal}. Using this, we find,
\begin{align}
\label{justthis}
G_{\rm obs}(f,\x,P_e^{\rm ex})&=\max_{k\in \{+1,-1\}} P_e^{\rm ex}(f(\a)=k|\x)\nonumber\\
&=|P_e^{\rm ex}(f(\a)=h_N|\x)-1/2|+1/2\nonumber\\
&=P_e^{\rm ex}(f(\a)=h_N|\x),
\end{align}
for every $e$.
Therefore, 
\begin{align}
\label{onemoreeqn}
G_{\rm int}(f,\x,P_{\rm obs})&=\max_{\{p(e|\x),P_e^{\rm ex}\}}  \sum_{e} p(e|\x) G_{\rm obs}(f,\x,P_e^{\rm ex}) \nonumber\\
&= \max_{\{p(e|\x),P_e^{\rm ex}\}}  \sum_{e} p(e|\x) P_e^{\rm ex}(f(\a)=h_N|\x) \nonumber \\
&= P_{\rm obs}(f(\a)=h_N|\x),
\end{align}
where the last equality follows from the constraint $\sum_e
p(e|\x) P_e(\a|\x)=P_{\rm obs}(\a|\x)$. On the other hand the
observed randomness for $f$ is, $G_{\rm obs}(f,\x,P_{\rm
obs})=P_{\rm obs}(f(\a)=h_N|\x)$. $\Box$

The previous technical results are valid for any non-signalling
distribution maximally violating the Mermin inequality. For odd
$N$ this maximal violation can be attained by a unique quantum
distribution, denoted by $P_{{\rm ghz}}(\a|\x)$, resulting from
measurements on a Greenberger-Horne-Zeilinger (GHZ) state. When
applying Theorem~\ref{Theorem1principal} to this distribution, one
gets

\textbf{Main result:} Let $P_{{\rm ghz}}(\a|\x)$ be the
$N$-partite (odd $N$) quantum probability probability distribution
attaining the maximal violation of the Mermin inequality. The intrinsic and observed randomness of $f$ for a Mermin input
satisfy

\be G_{\rm int/obs}(f,\x,P_{\rm ghz})= \frac{1}{2} + \frac{1}{2^{(N+1)/2}}
\ee

This follows straightforwardly from Theorem
\ref{Theorem1principal}, since $P_{\rm ghz}(\a|\x) = 1/2^{N-1}$
for outcomes $\a$ with an even number of results equal to $-1$ and
for those measurements
appearing in the Mermin inequality. 

It is important to remark that $f(\a|\x_m)$ approaches a
perfect random bit exponentially with the number of parties. In
fact, this bit defines a process in which full randomness
amplification takes place. Yet, it is not a complete protocol as,
contrary to the existing proposal in~\cite{Gallego2012a}, no
estimation part is provided.

\textit{Discussion ---} We have identified a quantum process whose
observed randomness can be proven to be fully intrinsic. In other
words, for the considered process, quantum theory gives
predictions as accurate as any no-signalling theory, possibly
supra-quantum, can give. Our results hold under the minimal
assumptions: the validity of the no-signaling principle and an
arbitrary (but not complete) relaxation of the freedom of choice.
The latter is subtle and much attention in recent years has
focused on relaxing it in Bell experiments \cite{Kofler2006,
Hall2011, Hall2010, Barrett2010, Koh2012}.

Our work raises several questions. Our main motivation here has
been to understand the ultimate limits allowed by quantum theory
on intrinsic randomness and, thus, we have worked in a noise-less
regime. It is interesting to consider how would our results have
to be modified to encompass scenarios including noise and hence
amenable to experiments. The presence of noise modifies our
results from two different viewpoints. First, noise is due to lack
of control of the setup and, thus, a source of apparent
randomness, which immediately implies a gap between intrinsic and
observed randomness.

Second, in a noisy situation, it is impossible to arbitrarily
relax the freedom of choice assumption, quantified by $\delta$ in
Eq.~(\ref{relaxfchoice}). In fact, there is a tradeoff between the
amount of relaxation of this condition and the violation needed to
certify the presence of any intrinsic randomness. The reason
is that, for a sufficiently small value of $\delta$,
any correlations not attaining the maximal non-signalling
violation of a Bell inequality can be reproduced using purely
deterministic local strategies. It seems natural, in a practical
context, to extend the definition of intrinsic randomness by
considering bounded relaxations of the freedom of choice
assumption and non-maximal violations of Bell inequalities. These
investigations would constitute the strongest tests on the
completeness of quantum predictions, given that they would rely on
significantly more relaxed assumptions than any other quantum
experiment performed to date.

From a purely theoretical perspective, our results certify a
maximum of one bit of randomness for any system size. It would be
interesting to extend these analytical results to certify
randomness that scale with the number of parties. This could for
instance be accomplished with functions of increasing outcomes. In
a related context, it would also be interesting to explore whether
similar results are possible in a bipartite scenario or, on the
contrary, whether an asymptotic number of parties is necessary for
full randomness amplification.

\begin{acknowledgements}
We acknowledge support from the ERC Starting Grant PERCENT, the EU Projects Q-Essence and QCS, the Spanish MICIIN through a Juan de la Cierva grant and the Spanish FPI grant,  an FI Grant of the Generalitat de Catalunya and projects FIS2010-14830, Explora-Intrinqra, CHIST-ERA DIQIP.
\end{acknowledgements}

\bibliographystyle{unsrt}    
\bibliography{thesis_bib}

\begin{widetext}\bigskip

\section{Appendix}

\appendix

Here we prove the principal theorem of the main text. It is
basically a generalization of the the proof for $N=3$. We would
like to prove that the function $f$ defined in the main text,
satisfies the property:
\begin{equation}\label{maineq}
    P(f(\a)=h_N|\mathbf{x_m})\geq 1/2
\end{equation}
for any $N$-partite distribution (odd $N$) that maximally violates
the Mermin inequality. As in the tripartite case, in order to
prove the result we (I) express condition \eqref{maineq} in terms
of some correlators and (II) use positivity conditions from the
swapped input to prove the inequality. 

An $N$-partite no-signalling probability distribution  $P(\a|\x)$
with inputs $\x\in\{0,1\}^N$ and outputs $\a\in\{+1,-1\}^N$  can
be parameterized in terms of correlators as,

\begin{equation}
\label{ProbDistrCorr}
P(\a|\x)=\frac{1}{2^N} \left(1+\sum_{i=1}^N a_i\la x_i\ra+ \sum_{i<j} a_i
a_j\la x_i x_j\ra+\nonumber\\
\sum_{i<j<k} a_i a_j a_k\la x_i x_j
x_k\ra+
\dots+a_1a_2\dots a_N\la x_1 x_2\dots x_N\ra\right).
\end{equation}
Restricting $P(\mathbf{a}|\mathbf{x})$ to those maximally
violating the $N$-partite Mermin inequality is equivalent to
requiring all correlators of input strings of odd parity to take
their extremal values. Namely, we have, \be \label{MerminCond} \la
x_1x_2\dots x_N\ra =(-1)^{(-1+\sum_{i=1}^N x_i)/2}, \ee for all
$N$-point correlators satisfying $\sum_{i=0}^N x_i=1 \; \text{mod}
\;2$. For instance, $\la 0,0,\dots,1\ra=1$ and similarly for all
permutations. Also, $\la 0,0,\dots,0,1,1,1\ra=-1$ as well as for
for all permutations, etc. In the following we will use the
notation $\la .\ra_k$ to denote a $k$-point correlator. The input
combination used to extract randomness is a generalization of the
tripartite case and denoted by $\mathbf{x_m}=(0,0,\ldots,0,1)$.
The corresponding $N$-point correlator  satisfies $\langle
0,0,\ldots, 0,1 \rangle=1$ for all $N$. The latter implies two
useful relations:
\begin{itemize}

\item[1.] Half the total outcomes vanish. In particular these are the terms for which the product of outcomes is $-1$ \ie $P(\prod_{i=1}^N a_i=-1|\mathbf{x_m})=0$.

\item[2.] $\la .\ra_{N-k}=\la .\ra_k$ for all $1\leq k\leq (N-1)/2$ where the correlators $\la .\ra_{N-k}$ and $\la .\ra_k$ are complementary in the input $\mathbf{x_m}$.
\end{itemize}
One can use these in Eqn.~\ref{ProbDistrCorr} to express
$P(\mathbf{a}|\mathbf{x_m})$ in terms of only the first
$(N-1)/2$-point correlators as,
\begin{equation}
\label{alpha}
P(\mathbf{a}|\mathbf{x_m}) =
\frac{1}{2^{N-1}} \left(1 +\sum a_i\la x_i\ra+\sum a_i
a_j\la x_i x_j\ra \\
 +\cdots
 +\sum a_ia_j\cdots a_p \la x_i x_j\cdots x_p\ra_{(N-1)/2}\right).
\end{equation}
where $a_1\cdot a_2\cdot a_3\dots a_N=+1$ since
$P(\a|\mathbf{x_m})=0$ when $a_1\cdot a_2\cdot a_3\dots a_N=-1$.


\section{Expressing the inequality in terms of correlators}

As mentioned, our first goal is to express Eq.\eqref{maineq} as a
function of some correlators. Let us recall the function we use in
our main theorem, 
\begin{equation}
\label{fdefined}
\begin{aligned}
f(\a)= \left\{
\begin{array}{c l}
+1~~ & n_-(\a)=(4j+2) ;~~{\rm with~} j\in \{0,1,2,\ldots\} \\
\\
-1~ & \text{otherwise}
\end{array}
\right .
\end{aligned}
\end{equation}
where $n_-(\a)$ denotes the number of results in $\a$ that are
equal to $-1$.

It turns out that the quantity (Eq. \eqref{maineq}) we would like to calculate, namely, $P(f(\a)=h_N|\x_m)-1/2$ can be equivalently expressed as $h_N\cdot(P(f(\a)=+1|\x_m)-1/2)$. The latter form is convenient since the function only takes value $+1$ for all $N$.

We proceed to express the latter in terms of correlators (as in the proof for three parties in the main text),
\begin{equation}
\label{compactP} \left(h_N \cdot P(f(\a)=+1|\x_m)-1/2\right)
=2^{-(N-1)} \boldsymbol\alpha' \cdot {\bf c} ,
\end{equation}
where
\begin{equation}
\label{compactP}
\begin{aligned}
& \boldsymbol\alpha'=h_N\cdot (\alpha_0-2^{N-2},\alpha_1,\alpha_2,\ldots,\alpha_{(N-1)/2}) \\
& {\bf c}=\left (1,\sum_{\mathcal{S}^1} \la
.\ra_1,\sum_{\mathcal{S}^2} \la
.\ra_2,\ldots,\sum_{\mathcal{S}^{(N-1)/2}} \la .
\ra_{(N-1)/2}\right)
\end{aligned}
\end{equation}
Note that, since the function $f$ symmetric under permutations,
the vector ${\bf c}$ consists of the different sums of all
$k$-point correlators, denoted by ${\mathcal S}^k$, where $k$
ranges from $0$ to $(N-1)/2$ because of Eq.~\eqref{alpha}. The
vector $\boldsymbol\alpha'$ is the vector of coefficients for each
sum of correlators. Our next goal is to compute this vector.

Recall that function $f$ is such that $f(\a)=+1$ if $n_{-}(\a)=4j+2$ for any $j\in \mathbb{N}\cup \{0\}$. By inspection, the
explicit values of $\alpha_i$ can be written as \be
\label{explicit} \alpha_i=\sum_{r=0}^{i} (-1)^r \binom{i}{r}
\sum_{j\geq 0} \binom{n-i}{4j+2-r}. \ee For example,
$\alpha_0=\sum_{j\geq 0} \binom{n}{4j+2}$ as one would expect
since $\alpha_0$ simply counts the total number of terms
$P(\a|\x_m)$ being summed to obtain $P(f(\a)=+1|\x_m)$.

Making use of the closed formula $\sum_{j\geq 0}
\binom{n}{rj+a}=\frac{1}{r}\sum_{k=0}^{r-1}\omega^{-ka}(1+\omega^k)^n$
~\cite{art}, where $\omega=e^{i2\pi/r}$ is the $r^{{\rm
th}}$ root of unity, we can simplify the second sum appearing in
Eq.~\ref{explicit}. Finally we recall that the phase $h_N$ was defined (in the main text) to be $h_N=\sqrt{2}\cos{(N+4)\pi/4}$.
Putting all this together and performing the first sum in
Eq.~(\ref{explicit}) gives us, \be \label{all_alpha}
\alpha_i'=2^{\frac{N-3}{2}}
\left(-2\cos{\frac{(N-2i)\pi}{4}}\cos{\frac{(N+4)\pi}{4}}\right)
\ee Notice that the term in the parenthesis is a phase taking
values in the set $\{+1,-1\}$ since $N$ is odd while the amplitude
is independent of $N$. Thus, we can simplify Eqn.
~(\ref{all_alpha}) for even and odd values of $i$ as,

\begin{equation}
\label{alpha_full}
\begin{aligned}
\alpha_i' = \left\{
\begin{array}{c l}
2^{(N-3)/2}(-1)^{\frac{N-i}{2}} & \text{$i$ odd} \\
\\
2^{(N-3)/2}(-1)^{\frac{i}{2}} & \text{$i$ even}
\end{array}
\right .
\end{aligned}
\end{equation}
Thus, to prove that $f$ possesses the property
$h_N\cdot (P(f(\a)=+1|\mathbf{x_m})-1/2)\geq 0$ necessary to proving the
main theorem is equivalent to proving \be \boldsymbol\alpha' \cdot
{\bf c} \geq 0, \ee for $\mathbf{c}$ as defined in
Eqn.~(\ref{compactP}) and for the values of $\boldsymbol\alpha'$
given by Eqn.~(\ref{alpha_full}). This is the task of the
following section, where we show that it follows from
positivity constraints on $P(\mathbf{a}|\mathbf{x})$.

\vspace{0.1cm}
\section{proving the inequality from positivity constraints}
We
show that  positivity conditions derived from the swapped input
$\mathbf{\bar{x}_m}=(1,1,\dots,1,0)$ may be used to show
$\boldsymbol\alpha' \cdot {\bf c} \geq 0$. Notice that the
components of $\mathbf{\bar{x}_m}$ and $\mathbf{x_m}$ are opposite, ie.
$\{\mathbf{\bar{x}_m}\}_i =\{\mathbf{x_m}\}_i\oplus 1$ for all $i$. In the
following we will repeatedly use the Mermin conditions of
Eqn.~\eqref{MerminCond}.

We start by summing the positivity conditions $P(+++\dots +-|\mathbf{\bar{x}_m})\geq 0$ and $P(---\dots -+|\mathbf{\bar{x}_m})\geq 0$. Using Eqn.~\eqref{ProbDistrCorr}, one can easily see that upon summing, all $k$-point correlators for \textit{odd} $k$ are cancelled out since these are multiplied by coefficients (products of $a_i$s) that appear with opposite signs in the two positivity expressions. In contrast, $k$-point correlators for \textit{even} $k$ add up since they are multiplied by coefficients that appear with the same sign in the two expressions. For example, $N$ being odd, the full correlator always cancels out while the $(N-1)$-point correlators always appear.

This leaves us with an expression containing only the even-body correlators,
\begin{equation} \label{sum}
1+\sum_{i<j} a_i a_j\la x_i x_j\ra
 +\sum_{i<j<k<l} a_i a_j a_k a_l\la x_i x_j
x_k x_l\ra +\\
 \dots +\sum a_i\dots a_p\underbrace{\la x_i\dots x_p\ra}_{(N-1)\text{-pt. corr}} \geq 0.
\end{equation}
Note once again, that this inequality is derived from the so-called swapped input $\mathbf{\bar{x}_m}$. We aim to cast it in a form that  can be compared directly with Eqn.~\eqref{compactP}, which comes from the chosen Mermin input $\mathbf{x_m}$. To this end, we need to convert Eqn. \eqref{sum} to an expression of the form,
\begin{equation}
\label{betai}
(\beta_0,\beta_1,\dots \beta_{(N-1)/2}).\\
 \left(1,\sum \la .\ra_1,\dots,\sum \la .\ra_{(N-1)/2}\right) \geq 0
\end{equation}

We first highlight the similarities and differences between the two preceding expressions, namely, the one we have \ie Eqn. \eqref{sum} and the one we want, \ie Eqn. \eqref{betai}. Each contains $(N-1)/2$ distinct classes of terms. However the former contains only even $k$-point correlators for $k=2$ to  $(N-1)$ while the latter contains all terms from $k=1$ to $(N-1)/2$. Thus, terms of Eq. \eqref{sum} must be mapped to ones in Eqn. \eqref{betai}. Moreover, since the point of making this mapping is to finally compare with Eqn. \eqref{compactP}, we also note that the correlators appearing in Eqn. \eqref{sum} are locally swapped relative to those appearing in Eqn. \eqref{compactP}. Thus, our mapping must also convert correlators of the swapped input into those corresponding to the chosen input.

We demonstrate next that one may indeed transform the inequality \eqref{sum} into the inequality \eqref{betai} satisfying both the demands above. To this end, all the \textit{even} $k$-point correlators (for $k \geq
\frac{N-1}{2}$) appearing in Eqn. \eqref{sum} are mapped to odd ($N-k$)-point correlators in Eqn.~\eqref{betai}. Likewise, all
the even $k$-point correlators (for $k <
\frac{N-1}{2}$) of the swapped input appearing in Eqn. \eqref{sum} are mapped to the corresponding $k$-point correlators of the chosen input
in Eqn. \eqref{betai}.

These mappings make systematic use of the Mermin conditions Eqn.~\eqref{MerminCond} and are made explicit in the following section.
\vspace{0.1cm}

\subsection{Even-point correlators} Consider a $2k$-point correlator where $2k\leq (N-1)/2$. The correlators are of two forms and we show how they are transformed in each case:
\begin{itemize}
\item $\la 11\dots 1\ra_{2k}$. We would like to map this to the correlator $\la 00\dots 0\ra_{2k}$ appearing in $\mathbf{x_m}$. We achieve the mapping by completing each to the corresponding Mermin full-correlators $\la \underbrace{11\dots 1}_{2k}\underbrace{100\dots 0}_{(N-2k)}\ra_N = (-1)^k$ and $\la \underbrace{00\dots 0}_{2k}\underbrace{100\dots 0}_{(N-2k)}\ra_N=(-1)^0=1$. From the signs, we have the relation, $\la 11\dots 1\ra_{2k} = (-1)^k\la 00\dots 0\ra_{2k}$

\item $\la 11\dots 10\ra_{2k}$, which we would like to map to $\la 00\dots 01\ra_{2k}$. Using the same ideas we get $\la \underbrace{11\dots 10}_{2k}\underbrace{110\dots 0}_{(N-2k)}\ra_N = (-1)^k$ and $\la \underbrace{00\dots 01}_{2k}\underbrace{110\dots 0}_{(N-2k)}\ra_N=(-1)^1=-1$. Thus, giving us the relation $\la 11\dots 10\ra_{2k} = (-1)^{k+1}\la 00\dots 01\ra_{2k}$.
\end{itemize}
By inspection one can write the relationship
\begin{equation*}
\underbrace{a_1a_2\dots a_{2k}}_{\rm even}\underbrace{\la x_1x_2\dots x_{2k}\ra}_{\text{cor in $\mathbf{\bar{x}_m}$}}=(-1)^k\underbrace{\la x_1x_2\dots x_{2k}\ra}_{\text{cor in $\mathbf{x_m}$}(desired)}
\end{equation*}
for correlators of either form discussed above on multiplying with their corresponding coefficients. Since we have finally converted to the desired correlators of the chosen input $\mathbf{\bar{x}}$, we can read off $\beta_i$ as the corresponding phase. Thus, $\beta_i = (-1)^{i/2}$ for even $i$.
\vspace{0.1cm}

\subsection{Odd-point correlators} Consider now a $2k$-point correlator where $2k\geq (N-1)/2$. The correlators are again of two forms and may be transformed to the required ($N-2k$)-point correlators in each case. The only difference from before is that the two correlators are now complementary to each other in the swapped input.
%
Since the details are similar, we simply state the final result $\beta_i=(-1)^{(N-i)/2}$ for odd $i$.

The final expression thus reads,
\begin{equation}
\label{betafull}
\begin{aligned}
\beta_i = \left\{
\begin{array}{c l}
(-1)^{\frac{N-i}{2}} & \text{$i$ odd} \\
\\
(-1)^{\frac{i}{2}} & \text{$i$ even}
\end{array}
\right .
\end{aligned}
\end{equation}
Thus, the values of $\beta$ given in Eqs. (\ref{betafull}) exactly match the ones for $\alpha_i'$ (up to the constant factor) given in Eqn. \ref{alpha_full}. Together with the correlators matching those in $\mathbf{c}$, it proves that $f$ satisfies the required $\boldsymbol\alpha' \cdot \mathbf{c}\geq 0$ and hence the full result.

\section{Proof that all distributions in decomposition maximally violate the Mermin inequality}
We end by proving the claim made in the main text that if an observed probability distribution $P_{\rm obs}(\a|\x)$ violates maximally and algebraically the corresponding Mermin inequality, all the no-signaling components $P_e^{\rm ex}(\a|\x)$ present in its preparation must also algebraically violate the inequality.

We recall that the decomposition appears in the definition of intrinsic randomness given by,
\begin{align}
& G_{\rm int}(g,\x,P_{\rm obs})= \max_{\{p(e|\x),P_e^{\rm ex}\} } \sum_{e} p(e|\x) G_{\rm obs}(g,\x,P_e^{\rm ex})\nonumber \\
 &\text{subject to:} \nonumber\\
&\sum_e p(e|\x) P_e^{\rm ex}(\a|\x)=P_{\rm obs}(\a|\x)  \label{sumupto} \\
&p(\x|e)\geq \delta \;\; {\rm with}~\delta>0 \;\; \forall\; \x,e
\end{align}

Since $P_{\rm obs}$ algebraically violates the Mermin inequality, this definition imposes stringent conditions on the correlators of $P_{\rm obs}$ satisfying the Mermin condition \eqref{MerminCond}, namely that,
\begin{equation}
\label{constraint_p}
\la x_1\ldots x_N \ra_{P_{\rm obs}} = \pm 1 = \sum_e p(e|x_1,\ldots,x_N)\; \la x_1\ldots x_N \ra_{P_e^{\rm ex}}
\end{equation}
where by normalization $\sum_e p(e|x_1,\ldots,x_N) =+1$ and $-1\leq\la x_1\ldots x_N \ra_{P_e^{\rm ex}}\leq+1$. Note that condition $p(\x|e)\geq \delta~{\rm for~all}~\x,e$ for $\delta>0$ can be inverted using the Bayes' rule to obtain $p(e|\x)> 0~{\rm for~all}~\x,e$. Now is clear by convexity that the condition $p(\x|e)\geq \delta$ (denying \textit{absolute} relaxation of freedom of choice) implies that all the correlator $\la x_1\ldots x_N\ra_{P_e^{\rm ex}}$ appearing in the Mermin inequality must also necessarily satisfy $\la x_1\dots x_N \ra_{P_e^{\rm ex}}=\pm 1$ for all $e$ thus maximally violating the Mermin inequality. In fact it is also clear that this constraint on $p(\x|e)$ {is strictly necessary} to ensure that the decomposition correlations satisfy maximal Mermin violation. To see this, suppose $p(\x|e_0)=0$, then the corresponding $\la x_1\dots x_N \ra_{P_{e_0}^{\rm ex}}$ is fully unconstrained while satisfying Eq. \eqref{constraint_p}.

\end{widetext}

\end{document}